\documentstyle[12pt]{article}
\def\<{{\langle}}
\def\>{{\rangle}}
\def\bra{{\langle}}
\def\ket{{\rangle}}
\def\ie{{\it i.e.}}
\def\be{\begin{equation}}
\def\ee{\end{equation}}
\def\ba{\begin{eqnarray}}
\def\ea{\end{eqnarray}}
\def\mref#1{Eq.(\ref{Eq:#1})}
\def\mreff#1{Fig.\ref{Eq:#1}}
\def\mreft#1{Table\ref{Eq:#1}}
\def\mlab#1{\label{Eq:#1}}
\def\mlabf#1{\label{Eq:#1}}
\def\mlabt#1{\label{Eq:#1}}
\def\half{\frac{1}{2}}
\def\to{\rightarrow}
\def\nn{\nonumber\\}
\def\sk{\vskip 1cm}
\def\skk{\vskip 3mm}
\def\mat#1#2#3{<{#1}\vert{#2}\vert{#3}> }
\def\etal{{\it et al.}}
\def\etc{{\it etc.~}}

\def\PR#1#2#3 {{\it Phys. Rev. }{\bf D#1} #2 {(#3)} }
\def\PRL#1#2#3 {{\it Phys. Rev. Lett. }{\bf #1} #2 {(#3)} }
\def\PL#1#2#3 {{\it Phys. Lett. }{\bf #1} #2 {(#3)}  }
\def\AP#1#2#3 {{\it Ann, Phys. }{\bf #1} #2 {(#3)} }
\def\ZP#1#2#3 {{\it Z. Phys. }{\bf #1} #2 {(#3)} }
\def\NP#1#2#3 {{\it Nucl. Phys. }{\bf #1} #2 {(#3)}  }
\def\MPL#1#2#3 {{\it Mod. Phys. Lett.}{\bf #1} #2 {(#3)}  }
\def\NC#1#2#3 {{\it Nuov. Cimm. }{\bf #1} #2 {(#3)}  }
\def\PREP#1#2#3 {{\it Phys. Report }{\bf #1} #2 {(#3)}  }
\def\PROG#1#2#3 {{\it Prog. Theor. Phys. }{\bf #1} #2 {(#3)}   }
\def\SOV#1#2#3{{\it Sov. J. Nucl. Phys. }{\bf #1} #2 {(#3)}   }
\def\JETP#1#2#3{{\it JETP}{\bf #1} #2 {(#3)}   }
\def\RMP#1#2#3{{\it Rev. Mod. Phys.}{\bf #1} #2 {(#3)}   }
\def\mat#1#2#3{<{#1}\vert{#2}\vert{#3}> }
\def\phiout{{\phi_a^{out}}}
\def\phiin{\phi_a^{in}}
\def\psiout{\psi_a^{out}}
\def\psiin{\psi_a^{in}}
\def\vin{\phi_{a\mu}^{in}}
\def\vout{\phi_{a\mu}^{out}}
\def\ofx{{(x)}}
\def\op{{\bf P}}
\def\oc{{\bf C}}
\def\ot{{\bf T}}
\def\cp{{\bf CP}}
\def\cpt{{\bf CPT}}
\def\vecr{{\vec r}}
\def\vecn{{\vec {\nabla}}}
\def\vecA{{\vec A}}
\def\psibar{\overline\psi}
\def\outin{{{out}\choose{in}}}
\def\inout{{{in}\choose{out}}}

\def\vr{\vec x}
\def\itt{{\it T}}
\def\b{{\bf b}}
\def\a{{\bf a}}
\def\d{{\bf d}}

\def\alphadot{{\dot\alpha}}
\def\betadot{{\dot\beta}}
\def\gammadot{{\dot\gamma}}

\def\N{\sqrt{{{E+mc^2}\over{2mc^2}}}}
\def\F#1{{{#1}\over{E+mc^2}}}
\def\itemm{\hangindent\parindent\textindent}
\def\noi{\noindent}
\def\onehead#1{\vskip1pc\leftline{\bf #1}}
\def\twohead#1{\vskip1pc\leftline{\bf #1}}
\def\ts{\thinspace}

\def\sq2{{1\over{\sqrt{2}}}}
\def\omegaar{{\vec{\omega}}}
\def\kbar{\overline K}
\def\Pbar{\overline P}
\def\Abar{\overline A}
\def\dbar{\overline d}
\def\ubar{\overline u}
\def\sbar{\overline s}
\def\bbar{\overline B}
\def\gbar{\overline g}
\def\pbar{\overline P}
\def\qbar{\overline q}
\def\vecr{\vec r}

\def\dm{\Delta m}
\def\ss{(1+s^2)}
\def\cdmp{cos\Delta m(t_1+t_2)}
\def\cdm{cos\Delta m(t_1-t_2)}

\def\g5{\gamma_5}
\def\gm{(1-\gamma_5)}
\def\gp{(1+\gamma_5)}

\def\mvec#1{\vec{#1}\,}
\def\pslash{\mbox{/\llap p}}
\def\slash#1{\mbox{/\llap #1}}

\def\msmall#1{\mbox{\rm \small #1}}
\renewcommand{\topfraction}{1.0}
\renewcommand{\bottomfraction}{1.0}
\renewcommand{\textfraction}{0.0}
\newcommand{\matel}[3]{\langle #1|#2|#3\rangle}
\newcommand{\hscale}{\mu\ind{hadr}}
\newcommand{\aver}[1]{\langle #1\rangle} 
\renewcommand{\Im}{\mbox{Im}\,}
\renewcommand{\Re}{\mbox{Re}\,}
\newcommand{\GeV}{\,\mbox{GeV}}
\newcommand{\MeV}{\,\mbox{MeV}}
\newcommand{\BR}{\,\mbox{BR}}

\begin{document}
\renewcommand{\thefootnote}{\fnsymbol{footnote}}

%\twocolumn[

\begin{flushright}
UND-HEP-99-BIG\hspace*{.2em}04\\
DPNU-99-13\\
hep-ph/9904484 \\ 
REVISED \\
%version 4.0 \\
\end{flushright}
%\today\\
%hep-ph/9904484\\
%Version 3.0 \\ 
%\end{flushright}
%\vspace{.3cm}
\begin{center} \Large 
{\bf On Limitations of \ot~Invariance in $K$ Decays}
\end{center}
\vspace*{.3cm}
\begin{center} {\Large 
I. I. Bigi $^{a}$, A. I. Sanda $^{b}$}\\
\vspace{.4cm}
{\normalsize 
$^a${\it Physics Dept.,
Univ. of Notre Dame du
Lac, Notre Dame, IN 46556, U.S.A.}\\
$^b$ {\it  Physics Dept., Nagoya University, 
Nagoya 464-01,Japan}}
\\
\vspace{.3cm}
e-mail addresses:\\
{\it bigi@undhep.hep.nd.edu, sanda@eken.phys.nagoya-u.ac.jp} 
\vspace*{.4cm}

{\Large{\bf Abstract}}\\
\end{center} 
Data from the CPLEAR collaboration coupled with the 
assumption that the Bell-Steinberger relation holds have provided  
{\em direct} evidence for \ot~violation. In this note we 
investigate what we can say about \ot~violation {\em without}  
such an assumption.

We show that  both the modulus and 
the phase of $\eta _{+-}$ can be reproduced with 
\ot~{\em invariant} dynamics through finetuning 
\cpt~breaking. The large 
\ot~odd correlation observed by the KTeV collaboration 
in $K_L \to \pi ^+\pi ^- e^+ e^-$ thus  
does not yield direct evidence for \ot~violation. In such a world 
the phase of $\frac{\epsilon'}{\epsilon}$ is 
$\delta_2-\delta_0-\phi_{SW}\sim -(85.5\pm 4)^\circ$. Also,
$K^{\pm} \to \pi ^{\pm} \pi ^0$ could exhibit a 
\cpt~asymmetry of up to few$\times 10^{-3}$ 
without upsetting any known 
bound. 
\vspace*{.2cm}
\vfill
\vskip 3mm

%]
%%%%%%%%%%%%
%%%%%%%%%%
\tableofcontents 
%%%%%%%%%%%%%% 
\section{Introduction}  
%%%%%%%%%%%%

\cpt~symmetry has an impressive {\em theoretical} pedigree 
as an almost inescapable consequence of Lorentz invariant 
quantum field theories. Observation of a \cp~asymmetry 
is therefore usually seen as tantamount to the 
discovery of a violation of time reversal invariance 
\ot . However the {\em experimental} verification 
of \cpt~invariance is much less impressive.   
Furthermore the 
emergence of superstring theories has opened -- by their 
fundamentally non-local structure -- a 
{\em theoretical} backdoor through which 
\cpt~breaking might slip in. 
This asks for carefully analysing the 
empirical basis of \cpt~invariance and the 
degree to which an observable can establish 
\ot~violation {\em directly}, i.e.  without invoking 
the \cpt~theorem\cite{ellis}.    
In addressing this issue, we will rely on as few 
other theoretical principles as possible: 
since we view the observation of \cpt~violation 
as a rather exotic possibility, we 
believe we should accept other theoretical restrictions very  
reluctantly only.  

Data from the CPLEAR collaboration have provided  
{\em direct} evidence for \ot~violation\cite{CPL}. In this note we want 
to address the following questions: 
\begin{itemize} 
\item 
To which degree and in which sectors of $\Delta S \neq 0$ 
dynamics 
is \ot~violated? 
\item 
How accurately is the validity of \cpt~invariance established 
{\em experimentally}? 
\item 
Which conclusions can be drawn {\em without} invoking the 
Bell-Steinberger relation. 
\item 
Which is the most promising -- or the least hopeless -- 
observable for finding \cpt~violations in kaon decays? 
\end{itemize}  

The reader might wonder why we are insisting on analyzing \ot~
symmetry without assuming the Bell-Steinberger relation. After all,
it is viewed as just a consequence of unitarity.
Yet the following has to be kept in mind: when   
contemplating the possibility of \cpt~violation 
-- a quite remote and exotic scenario --  we should 
not consider the Bell-Steinberger relation sacrosanct. 
The latter is based on the assumption that all relevant 
decay channels are known. 
Since the major branching fractions have been 
measured with at best an error of 1\%, some yet 
undetermined decay mode  
with a branching fraction of $10^{-3}$ can easily 
be hidden \cite{kabir}.
We are 
{\em not} arguing that this is a likely scenario -- it is 
certainly not! However we do not view it to be more exotic than 
\cpt~violation. Then it does not make a lot of sense to us to 
allow for the latter while forbidding the former.

The paper will be organized as follows: 
after briefly reviewing the formalism relevant 
for $K^0 - \bar K^0$ oscillations in 
Sect. \ref{FORMALISM} we list  
the direct evidence for \ot~being violated in Sect. \ref{T}; 
in Sect. \ref{CPT} we analyse the phases of 
$\eta _{+-}$ and $\eta _{00}$; 
after evaluating what can be learnt from 
$K_L \to \pi ^+\pi ^- e^+e^-$ in Sect. \ref{TODD}, we 
give our conclusions in Sect. \ref{SUMMARY}. 

%%%%%%%%%%%%%%
\section{Formalism 
\label{FORMALISM}}
%%%%%%%%%%%
To introduce our notation and make the paper self-contained 
we shall 
record here the standard formalism for the neutral 
$K$ meson system.

%%%%%%%%%%%%%%%%%%%%%
\subsection{$\Delta S=2$ Transitions}
%%%%%%%%%%%%%%%%%%

The time dependence of the state $\Psi$, which is a 
linear combination
of $K^0$ and $\overline K^0$, is given by
\be 
i \hbar \frac{\partial}{\partial t} \Psi(t) = {\cal H}\Psi(t) 
\; \; \; , \; \; \; 
\Psi (t) = \left( \matrix {K^0(t) \cr \overline K^0(t) \cr } \right). 
\mlab{Schroed2} 
\ee 
The $2 \times 2$ matrix ${\cal H}$ can be expressed 
through the 
identity and the Pauli matrices \cite{lee}
\be  
{\cal H} \equiv {\bf M} - \frac{i}{2} {\bf \Gamma} = 
E_1 {\bf \sigma}_1 + E_2 {\bf \sigma}_2 + 
E_3 {\bf \sigma}_3 - i D {\bf 1} .
\mlab{CPTMass} 
\ee 
with 
\ba
E_1 &=& \Re M_{12}-{i\over 2} \Re \Gamma_{12}\;  , \; \; 
E_2 = -\Im M_{12}+{i\over 2} \Im \Gamma_{12} 
\nn
E_3 &=& {1\over 2} (M_{11}-M_{22})-{i\over 4} 
(\Gamma_{11}-
\Gamma_{22})  , \;  
D = {i\over 2} (M_{11}+M_{22})+{1\over 4} 
(\Gamma_{11}+\Gamma_{22}).
\mlab{def2}
\ea
It is often convenient to use instead {\em complex} numbers 
$ E, \theta$, and $\phi$ defined by 
\ba
E_1=E\, {\rm sin}\theta \, {\rm cos}\phi ,~~ 
E_2&=&E\, {\rm sin}\theta \, {\rm sin}\phi,~~
E_3=E\, {\rm cos}\theta \nn 
E&=&\sqrt{E_1^2+E_2^2+E_3^2} 
\; . 
\mlab{6.27}
\ea
The mass eigenstates are given by 
\ba  
|K_S \rangle &=& p_1 |K^0\rangle + q_1 |\overline K^0\rangle \nn  
|K_L \rangle &=& p_2 |K^0\rangle - q_2 |\overline K^0\rangle 
\mlab{ES} 
\ea 
with the {\em convention} 
$\cp |K^0 \rangle = |\overline K^0 \rangle$ and  
\ba 
p_1 &=& N_1 {\rm cos}\frac{\theta}{2} , \;   
q_1 = N_1 e^{i \phi}{\rm sin}\frac{\theta}{2} \nn  
p_2 &=&  N_2 {\rm sin}\frac{\theta}{2}  ,  \;  
q_2 = N_2 e^{i \phi}{\rm cos}\frac{\theta}{2} \nn
N_1 &=& \frac{1}{\sqrt{|{\rm cos}\frac{\theta}{2}|^2 
+ |e^{i \phi}{\rm sin}\frac{\theta}{2}|^2}} \nn  
N_2 &=& \frac{1}{\sqrt{|{\rm sin}\frac{\theta}{2}|^2 
+ |e^{i \phi}{\rm cos}\frac{\theta}{2}|^2}}\; . 
\ea 

The discrete symmetries impose the following 
constraints: 
\ba
\cpt~{\rm or}~ \cp~{\rm invariance}~~
\Longrightarrow &&\cos\theta =0,
~~M_{11}=M_{22},~~\Gamma _{11}=\Gamma_{22}\nn 
\cp~{\rm or}~\ot~{\rm invariance}~~\Longrightarrow &&
\phi =0,~~ \Im M_{12}=0={\rm Im}\Gamma_{12}
\mlab{DS2CON}
\ea  

%%%%%%%%%%%%%%%%%%%%
\subsection{Nonleptonic Amplitudes}
%%%%%%%%%%%%%%%%%%%%%

We write for the amplitudes describing decays into 
final states with isospin $I$:  
\ba
 T(K^0 \to [\pi \pi ]_I) &=& A_I e^{i\delta _I},\nn 
T(\overline K^0 \to [\pi \pi ]_I) &=& \overline A_I e^{i\delta _I}
\ea
where the strong phases $\delta _I$ 
have been factored out and 
find: 
\ba
\cpt \; \; {\rm invariance} \; \;  
 \Longrightarrow && A_I = \overline A^*_I \nn 
\cp \; \; {\rm invariance} \; \;  
\Longrightarrow && A_I = \overline A_I \nn 
\ot \; \; {\rm invariance} \; \;  
 \Longrightarrow && A_I = A^*_I  
\ea  
The expressions for $\eta_{+-}$ and $\eta_{00}$
\ba 
\eta _{+-} &=&  \frac{1}{2} \left( \Delta _0 - 
\frac{1}{\sqrt{2}} \omega e^{i(\delta _2 - \delta _0)} 
(\Delta _0 - \Delta _2) \right),  \nn 
\eta _{00} &=& \frac{1}{2} \left( \Delta _0 +  
\sqrt{2} \omega e^{i(\delta _2 - \delta _0)} 
(\Delta _0 - \Delta _2) \right),  \nn 
\Delta _I &=& \frac{1}{2} \left( 1 - \frac{q_2}{p_2} 
\frac{\overline A_I}{A_I} \right) \; ,  \; 
|\omega | \equiv \left|\frac{A_2}{A_0}\right| \simeq \frac{1}{20},
\mlab{ETA} 
\ea 
are valid 
{\em irrespective} of \cpt~symmetry.

%%%%%%%%%%%%%%%%%%%%%%
\subsection{Semileptonic Amplitudes}
%%%%%%%%%%%%%%%%%
The general amplitudes for semileptonic $K$ decays 
can be expressed as follows:  
\ba
\bra l^+\nu \pi^-|{\cal H}_W|K^0\ket &=&F_l(1-y_l)\nn
\bra l^+\nu \pi^-|{\cal H}_W|\overline K^0\ket 
&=&x_l F_l(1-y_l)\nn
\bra l^-\overline\nu \pi^+|{\cal H}_W|K^0\ket &=&
\overline x_l^*F_l^*(1+y_l^*)\nn
\bra l^-\overline\nu \pi^+|{\cal H}_W|\overline K^0\ket 
&=&F_l^*(1+y_l^*).
\mlab{9.1000}
\ea
with the selection rules 
\begin{tabbing}
{\hskip 1cm}\=$\Delta S=\Delta Q$ rule:\hskip 2cm\= 
$x_l=\overline x_l=0$\\
\>\cp~invariance:  \>$x_l=\overline x_l^*;~~~F_l=F_l^*;
~~~y_l=- y_l^*$\\
~\> \ot~invariance:\>  ${\rm Im}~F={\rm Im}~y_l=
{\rm Im}~x_l={\rm Im}~\overline x_l=0$\\
~\> \cpt~invariance:\> $y_l=0,~x_l=\overline x_l$.
\end{tabbing}

%%%%%%%%%%%%%%%%%
\section{Direct Evidence for \ot~Violation 
\label{T}}
%%%%%%%%%%%%%

The so-called Kabir test\cite{kabir2} represents a quantity that 
probes \ot~violation without reference to 
\cpt~symmetry: 
\be 
A_{\ot} \equiv \frac{\Gamma (K^0 \to \overline K^0) - 
\Gamma (\overline K^0 \to  K^0)} 
{\Gamma (K^0 \to \overline K^0) +  
\Gamma (\overline K^0 \to  K^0)}
\mlab{KABIR} 
\ee 
A nonvanishing $A_\ot$ requires 
\be
M_{12}-\frac{i}{2}\Gamma_{12}\ne
M_{21}-\frac{i}{2}\Gamma_{21}.
\ee
which constitutes \cp~as well as \ot~ violation. 
Associated production flavor-tags the 
{\em initial} kaon. The flavor of the 
{\em final} kaon is inferred from semileptonic 
decays; i.e., we measure the 
\cp~asymmetry 
\be 
A_{\cp} \equiv  \frac{\Gamma(K\to l^-\nu\pi^+)-
\Gamma(\overline K\to l^+\nu\pi^-)}
{\Gamma(K\to l^-\nu\pi^+)+
\Gamma(\overline K\to l^+\nu\pi^-)} 
\mlab{1}
\ee 
Yet a violation of 
\cpt~invariance and/or of the $\Delta S = \Delta Q$ rule 
can produce an asymmetry in the latter 
-- $A_{\cp} \neq 0$ -- without one being 
present in the former -- $A_{\ot} =0$. 
These issues have to be tackled first. 
There is nothing new in our remarks on this subject; 
we add them for clarity and completeness.

%%%%%%%%%%%%%%
%\subsubsection{Bounds on \cpt~violating Parameters}
%%%%%%%%%%%%%

Analysing the asymmetries in 
$\Gamma(\overline K^0(t)\to l^+\nu K^-)$ vs. 
$\Gamma( K^0(t)\to l^- \bar \nu K^+)$ and 
$\Gamma(\overline K^0(t)\to l^-\bar \nu K^+)$ vs. 
$\Gamma( K^0(t)\to l^+ \nu K^-)$ 
for large times $t$ CPLEAR has found \cite{CPLEARCPT}
\be
\Re\cos\theta= (6.0\pm 6.6\pm 1.2)\times 10^{-4}.
\mlab{cp1}
\ee
>From the decay rate evolution they have inferred  
\ba
\Im\cos\theta= (-3.0\pm 4.6\pm 0.6)\times 10^{-2},\nn
\half\Re(x_l-\bar x_l)= (0.2\pm 1.3\pm 0.3)\times 10^{-2},\nn
\half\Im(x_l+\bar x_l)= (1.2\pm 2.2\pm 0.3)\times 10^{-2} \; . 
\mlab{CPTSLDATA} 
\ea
While there is no sign of \cpt~violation in any of these 
observables, the bounds of \mref{CPTSLDATA} are not overly  
restrictive.

%\be 
%A_\delta(t)=\frac{\Gamma(\overline K^0(t)\to l^+)-
%\Gamma(K^0(t)\to l^-)
%\alpha}{\Gamma(\overline K^0(t)\to l^+)+\Gamma(K^0(t)\to l^-)
%\alpha} + 
%\frac{\Gamma(\overline K^0(t)\to l^-)-\Gamma(K^0(t)\to l^+)
%\alpha}{\Gamma(\overline K^0(t)\to l^-)+\Gamma(K^0(t)\to l^+)
%\alpha}
%\ee 
%where $\alpha=|\frac{p_2}{q_2}|^2\xi$ with $\xi$ 
%denoting the ratio of 
%detection efficiencies for $K^+\pi ^-$ vs. 
%$K^-\pi ^+$ pairs. It is a neat feature of this 
%measurement that $\alpha$ can be determined
%from the data. 
%For large times $t$, $A_\delta(t)\to 4\Re\cos\theta$ and 

%%%%%%%%%%%%
%\subsubsection{Data on the Kabir Test}
%%%%%%%%%%%%%%
Another input is provided by the charge asymmetry 
in semileptonic $K_L$ decays for which the general 
expression reads as follows:  
\ba 
\delta _{\rm Lept} &=& 
\frac{\Gamma (K_L \to l^+ \nu \pi ^-) - 
\Gamma (K_L \to l^- \nu \pi ^+)}
{\Gamma (K_L \to l^+ \nu \pi ^-) + 
\Gamma (K_L \to l^- \nu \pi ^+)}\nn
&= &{\rm Im}\, \phi - {\rm Re \, cos}\, \theta - 
{\rm Re}\; (x_l - \overline x_l) - 2 {\rm Re}\, y_l.
\ea 
\cpt~violation, if it exists, is most likely to surface in $M_{12}$,  
which is of second order in the weak interactions. It is 
then natural to assume {\em semileptonic} decay amplitudes 
to conserve \cpt~, which is fully consistent with 
\mref{CPTSLDATA}, but not confirmed to the required level: 
\be
x_l-\bar x_l = 0,~~~~~~~ {\rm or}~~~~~~y_l= 0.
\ee
With this 
{\em assumption}, and from the data \cite{PDG} 
\be 
\delta _{\rm Lept} = (3.27 \pm 0.12) \times 10^{-3} .
\ee 
one obtains 
\be
{\rm Im}\, \phi - {\rm Re \, cos}\, 
\theta =(3.27 \pm 0.12) \times 10^{-3} 
\ee
and infers from \mref{cp1} 
\be
\Im\phi=(3.9\pm 0.7)\times 10^{-3},
\ee
showing that \ot~is violated in kaon dynamics.

This result can be stated more concisely as follows 
\cite{CPLEAR2}:
\be 
A_{T} \simeq A_{\cp } = 
(6.6 \pm 1.3\pm 1.0 ) \times 10^{-3}. 
\mlab{1}
\ee 

In order to get a result independent of the assumption 
that {\em direct semileptonic} kaon decays obey 
\cpt~symmetry,  
the CPLEAR collaboration has employed constraints from the 
Bell-Steinberger relation to deduce the bound \cite{CPL}
\be
\half{\rm Re}\; (x_l - \overline x_l) - {\rm Re}\, 
y_l =(-0.4\pm 0.6)\times 10^{-3} \; , 
\ee
which again is fully consistent with \cpt~invariance of the 
semileptonic decays. This results in establishing violation of
\ot~symmetry -- provided the assumption mentioned above is valid.

%%%%%%%%%%%%%%
\section{Phases of $\eta _{+-}$ \& $\eta _{00}$ 
and \cpt 
\label{CPT}}
%%%%%%%%%%%%%
%%%%%%%%%%%%%%%%%%%%%
\subsection{Basic Expressions}
%%%%%%%%%%%%%%%%%%%%%%%%%%5

Manipulating \mref{ETA} we obtain 
through ${\cal O}(\phi)$ and 
${\cal O}$(cos$\theta$)
\be   
|\eta _{+-}|
\frac{\Delta\Phi}{{\rm sin}\phi _{SW}}=
\left( \frac{M_{\overline K} - M_{K}}{2\Delta M} + 
R_{direct} \right) 
\mlab{CPTTEST} 
\ee
\ba 
\Delta\Phi &\equiv& 
\frac{2}{3} \phi _{+-} + \frac{1}{3}\phi _{00} 
- \phi _{SW} \nn 
R_{direct} &=& 
\half \Re~ r_A -\frac{ie^{-i\phi_{SW}}}{\sin\phi_{SW}}
\sum_{f\ne [2\pi]_0}\epsilon(f) \nn 
r_A &\equiv& \frac{\overline A_0}{A_0} - 1, 
\; \; \phi_{SW}\equiv 
{\rm tan}^{-1}\frac{2\Delta M}{\Delta \Gamma}\nn
 \epsilon(f) &=& e^{i\phi_{SW}}i\cos\phi_{SW}
\frac{\Im \Gamma_{12}(f)}
{\Delta \Gamma} \; . 
%\nonumber 
\ea 
Since \cpt~symmetry predicts $M_K=M_{\overline K}$
and $\Re~r_A={\cal O}(\xi_0^2)$, where $\xi_0=\arg~A_0$, 
it implies 
$|\Delta\Phi|=0$ to within the 
uncertainty given by $|\sum_{f\ne [2\pi]_0}\epsilon(f)|$; 
the latter sum thus represents the theoretical `noise'. 

%%%%%%%%%%%%%%%%%%%%%%%%
\subsection{Estimating $\sum \epsilon (f)$ 
%\cite{SCHUBERT}
}
%%%%%%%%%%%%%%%%%%%%%%%

The major kaon decay modes fall into two classes, namely 
flavor-{\em non}specific or flavor-specific channels. 
\begin{itemize}
\item 
With  
$A_f=\matel{f}{H_W}{K^0}$ and $\overline A_f=
\matel{f}{H_W}{\overline K^0}$, 
we have, to first order in \cp~violation,
\be
\Im \Gamma_{12}(f)=i\eta_f\Gamma(K\to f)\left( 1- 
\eta _f
\frac{\overline A_f}{A_f}\right) \; ,
\mlab{iden}
\ee
for \cp~eigenstates with eigenvalue $\eta _f$.
Im $\Gamma _{12}(f) \neq 0$ can hold only if 
$\overline A_f \neq \eta _f A_f$, i.e. if there is 
{\em direct} \cp~violation  in the channel $f$. 

Using data on $\epsilon ^{\prime}$, 
Br($K_{L,S} \to 3\pi$) and 
\be
{\rm Im}\, \eta _{+-0} =  
\left(-2 \pm 9~{+2\atop -1}\right)\times 10^{-3}\; , 
\; \; 
\Im\eta _{000}=0.07\pm 0.16  
\; \; \; \cite{3pi,BLOCH},
\mlab{Im410} 
\ee
where 
\be 
\eta_{+-0,000} \equiv 
\half \left( 1 + \frac{q_1}{p_1}
\frac{\bar A(\pi ^+ \pi ^- \pi ^0, 3\pi ^0)} 
{A(\pi ^+ \pi ^- \pi ^0, 3\pi ^0)}\right),   
\ee
we obtain 
\ba
|\epsilon(3\pi ^0)| &<& 1.1 \times 10^{-4}\nn
|\epsilon([2\pi ]_2)| &\simeq& 0.28 \times 10^{-6}\nn   
|\epsilon((\pi ^+ \pi^- \pi ^0 )_{\cp~-[+]})| &<&  
5\, [0.2]\times 10^{-6},  
\ea
\item 
Allowing for a violation of the 
$\Delta Q = \Delta S$ rule in semileptonic decays 
as expressed by 
$ 
x_l \equiv \frac{\matel{l^+ \nu \pi ^-}
{{\cal H}_W}{\overline K}}
{\matel{l^+ \nu \pi ^-}{{\cal H}_W}{K}}$,  
we find 
\be 
|\epsilon (\pi l \nu )| \leq 4 \times 10^{-7}. 
\ee 
\end{itemize} 
%%%%%%%%%%%%%%%%%%
\subsection{Quantifying \cpt~Tests}
%%%%%%%%%%%%%%%%%%%%%%%%%

With the measured values for the phases 
$\phi _{+-},~\phi _{00}- \phi _{+-}$,
and $\phi _{SW}$
we arrive at a result quite consistent with zero 
\cite{PDG,adler6,SCHUBERT}:
\be
\Delta\Phi=  
0.01^o \pm  
0.7^o |_{exp.} 
\pm 1.5^o |_{theor.} \; , 
\mlab{CPTTEST} 
\ee 
i.e., the phases $\phi _{+-}$ and $\phi _{00}$ agree with their 
\cpt~prescribed values to within $2^o$ .
\cpt~invariance is thus probed to about the 
$\delta \phi /\phi _{SW} \sim 
5\% $ level. 
The relationship between $\phi _{+-}$, $\phi _{00}$ on one 
side and $\phi _{SW}$ on the other is a truly 
meaningful gauge; yet the numerical accuracy of that test is 
not overwhelming. The theoretical error can be reduced significantly 
by making quite reasonable assumptions on \cp~violation; however, 
we refrain from doing so based on our belief that assuming 
observable \cpt~breaking is not very reasonable to start with.  

In \mref{CPTTEST}, the {\em theoretical} uncertainty 
$\sum_f\epsilon(f)$ provides the limiting factor for 
this test\cite{shabalin}; it is dominated by 
$K \to 3\pi ^0$. Future experiments could reduce the uncertainty 
by a factor of up to two \cite{BLOCH}.

Alternatively we can state 
\be
\frac{M_{\overline K} - M_K}{2\Delta M}  
+ \frac{1}{2}  {\rm Re}~ r_A 
= \left( 0.06 \pm 4.0|_{exp} 
\pm 9 |_{theor} 
 \right)  \times 10^{-5} .
\ee 
Yet  
$\Delta M$ does not provide a meaningful calibrator; for it 
arises basically unchanged  
even if \cp~were conserved while the latter would imply 
$M_{\overline K} - M_K = 0$ and $r_A=0$ irrespective of 
\cpt~breaking. 

The often quoted 
truly spectacular bound (for  $R_{direct} = 0$) 
\be 
\frac{M_{\overline K} - M_K}{ M_K}  = 
(0.08 \pm 5.3|_{exp}) \times 10^{-19} 
\ee   
definitely 
overstates the numerical degree to which 
\cpt~invariance has been probed. $M_K$ is not generated 
by weak interactions and thus cannot serve as a meaningful  yardstick.

In summary: while no hint has has found 
indicating a limitation to \cpt~symmetry, the {\em experimental} 
evidence for it is far from overwhelming: 
\begin{itemize}
\item 
Comparing the phases of $\eta _{+-}$ and $\eta _{00}$ with the 
superweak phase constitutes a meaningful test of 
\cpt~symmetry. Yet there is a `noise' level of about 
$2^o$ that cannot be reduced significantly \cite{BLOCH}. 
\item 
Relating the bound on the difference 
$| M_{\overline K} - M_K|$ to the 
kaon mass itself is extremely impressive numerically -- 
yet meaningless. 
\item 
When entertaining the idea of \cpt~violation, we should not 
limit our curiosity to a single quantity like 
$\Delta \Phi$ (or equivalently  
$M_{\overline K} - M_K$). 
\item
Finally, the reader should be reminded that \cpt~symmetry implies 
$\Delta\Phi\ll\phi_{SW}$ but the converse does not follow. 

\end{itemize}

%%%%%%%%%%%%%
\section{Consequences in a \ot~Conserving World \label{TODD}}
%%%%%%%%%%%%

%%%%%%%%%%%%%%%%%%%%
\subsection{Reproducing $\eta _{+-}$}
%%%%%%%%%%%%%%%%%%
Assuming nature to conserve \ot , 
which implies $\phi=0$, see \mref{DS2CON}, we have: 
%in \mref{eps2,CPTTEST}: 
%\be
%\epsilon=e^{i\phi_{SW}}\sin\phi_{SW}
%\frac{1}{\Delta M}\left(
%-\frac{i}{2}(M_{11}-M_{22})
%-\frac{1}{4}(\Gamma_{11}-\Gamma_{22})\right)
%-\half\Re r_A.
%\ee
\ba  
&&\frac{|\eta _{+-}|\Delta\Phi}{{\rm sin}\phi _{SW}}=
- \frac{M_{11} - M_{22}}{2\Delta M} 
+\half r_A  ,\nn
&&
\frac{|\eta_{+-}|}{\cos\phi_{SW}}=-\frac{\Gamma_{11}-
\Gamma_{22}}{4\Delta M}{\rm tg}\phi_{SW}
-\half r_A.\nn
&&\Re\cos\theta= - 
\frac{M_{11}-M_{22}}{\Delta M}\sin^2\phi_{SW}  
+\half\frac{\Gamma_{11}-\Gamma_{22}}{\Delta M}
\sin\phi_{SW}\cos\phi_{SW}.
\mlab{3}
\ea

%Therefore 
%we will not invoke the Bell-Steinberger relation in this 
%context; i.e. we will not impose \mref{BELL}! 
%In that case \mref{cv2} guarantees that 
%$\eta_{+-}\simeq|\eta_{+-}|e^{i\phi_{SW}}\neq 0$; 
%i.e., we predict
%the same value for $\langle A \rangle$ than in the 
%usual \ot~violating, \cpt~conserving scenario! 

Inserting the values of $\eta_{+-}$, $\phi_{SW}$ and \mref{cp1}  
%into Eqs.(\ref{cv2}) and \mref{3} 
we can solve for the three unknowns:
\ba 
\frac{M_{11}-M_{22}}{\Delta M}&\simeq& r_A\simeq 
(-3.9 \pm 0.7) \times 10^{-3} 
%(-1.7\pm 0.3)|\eta_{+-}| 
\nn 
\frac{\Gamma_{11}-\Gamma_{22}}{\Delta M}
&\simeq& (- 5.0 \mp 1.4) \times 10^{-3} . 
%(-2.3\pm 0.6)|\eta_{+-}|.
\mlab{cptviol}
\ea
The solution is very 
{\em unnatural} -- \mref{cptviol}, for example, 
requires cancellation 
between \cpt~violating $\Delta S=1$ and 2 amplitudes.
Yet however unnatural they may be, we must entertain 
this  possibility 
unless we can exclude it empirically.

As a side remark, we mention that
if we invoke the Bell-Steinberger relation in its usual 
form -- meaning that kaon decays are effectively saturated 
by the $K \to 2\pi , \, 3\pi , \, l \nu \pi$ channels, then 
we have an additional relation \cite{bloch2}:  
\be 
\half r_A \approx -\frac{\Gamma_{11} - \Gamma_{22}}{4\Delta M} \; ; 
\mlab{b1}
\ee 
i.e, \mref{3} then implies $\eta _{+-} \simeq 0$. 
This is not surprising since these known modes do 
not exhibit any sign of \cpt~violation. 
But, as we have remarked before, in testing 
\cpt~we want to stay away from invoking saturation 
by the known channels.
%%%%%%%%%%%%%%%%%%%%%%%%%%%%%%%%%%%%%%%%%
\subsection{$K\to\pi\pi$}
%%%%%%%%%%%%%%%%%%%%%%%%%%%%%%%%%
Where should such a large \cpt~violation show its face? 
Imposing 
$r_A \neq 0$ raises the prospects of unacceptably large 
direct \cp~violation in $K_L \to \pi \pi$. 
\mref{ETA} can be reexpressed as follows:  
\be
\epsilon \simeq  \frac{1}{\sqrt{1+(\frac{\Delta \Gamma}
{2\Delta M})^2}}e^{i\phi _{SW}}
\left( -\frac{{\rm Im}M_{12}}{\Delta M_K} + \xi _0\right) 
\mlab{6.3.12}
\ee 
\be
\epsilon ^{\prime} = 
\frac{1}{2\sqrt{2}}\omega e^{i(\delta _2 - \delta _0)} 
\frac{q_2}{p_2} 
\left( \frac{\overline A_0}{A_0} - \frac{\overline A_2}{A_2}\right) 
\mlab{6.107}
\ee
If \ot~is conserved, 
$\frac{q_2}{p_2}\left( \frac{\overline A_0}{A_0} - 
\frac{\overline A_2}{A_2}\right) $ is real
and \mref{6.107} then tells us \cite{PDG}
\be 
\arg\left(\frac{\epsilon'}{\epsilon}\right)
=\delta_2 - \delta_0 -\phi_{SW}\simeq - (85.5\pm 4)^\circ.
\mlab{RE0} 
\ee 
Therefore   
\ba 
{\rm Re} \frac{\epsilon ^{\prime}}{\epsilon} &\simeq& 
\cos (\delta _2 - \delta _0 - \phi _{SW}) \cdot  
\frac{|\omega|}{2\sqrt{2} |\eta _{+-}|} \cdot 
|\Delta _0 - \Delta _2| \nn 
&=& 0.035 \cdot \left( 0.087 ^{+0.061}_{-0.078} \right) \cdot 
\left| \frac{r_A^{\prime}}{r_A} - 1 \right| = 
\left( 3.0^{+2.2}_{-2.7}\right) \cdot 10^{-3}  
\cdot \left| \frac{r_A^{\prime}}{r_A} - 1 \right|   
\ea
where 
\be 
r_A^{\prime} \equiv \frac{\overline A_2}{A_2} - 1
\mlab{RP} 
\ee
Some remarkable features can be read off from this expression:
\begin{itemize}
\item 
For 
\be 
\delta _2 - \delta _0 - \phi _{SW} = 90 ^{\circ} 
\ee
which is still allowed by the data, one obtains 
\be 
{\rm Re} \frac{\epsilon ^{\prime}}{\epsilon} = 0 \; . 
\ee
As far as $K \to \pi \pi$ is concerned this amounts to a superweak 
scenario! 
\item 
The empirical landscape of \cp~violation has changed 
{\em qualitatively}: KTeV, confirming earlier observations of 
NA 31, has conclusively established the existence of 
direct \cp~violation \cite{KTEVPRL}:  
\be 
{\rm Re} \frac{\epsilon ^{\prime}}{\epsilon} = 
\left( 2.80 \pm 0.30 \pm 0.28 \right) \cdot 10^{-3} 
\ee
Including previous data and preliminary results from NA 48 one 
arrives at a world average of 
\be 
{\rm Re} \frac{\epsilon ^{\prime}}{\epsilon} = 
\left( 2.12 \pm 0.28 \right) \cdot 10^{-3} 
\ee 
This can be reproduced with a `canonical' $r^{\prime}_A = 0$, but only 
for a very narrow slice in the phase of $\epsilon
^{\prime}/\epsilon$, namely 
\be 
\delta _2 - \delta _0 - \phi _{SW} \simeq - (86.5 \pm 0.5)^{\circ} \;
. 
\mlab{SLICE}
\ee 
\item 
The dominant uncertainty here enters through the phase shifts 
$\delta _{0,2}$. If $\delta _2 - \delta _0 - \phi _{SW}$ falls outside
the range of \mref{SLICE}, then $r_A^{\prime} \neq 0$ is needed to
reproduce 
Re$(\epsilon ^{\prime}/\epsilon)$. As an illustration consider 
$ \delta _2 - \delta _0 - \phi _{SW} = 80^{\circ}$. In that case 
$1/2 \leq r_A^{\prime}/r_A \leq 5/6$ had to hold to obtain 
$1 \cdot 10^{-3}\leq {\rm Re}(\epsilon ^{\prime}/\epsilon )\leq 3 \cdot
10^{-3}$. Hence $r_A^{\prime} \sim - (2 \div 4) \cdot
10^{-3}$. More generally if 
\be 
\delta _2 - \delta _0 - \phi _{SW} \leq 83 ^{\circ} 
\ee
then the observed value of Re$(\epsilon ^{\prime}/\epsilon)$ would
imply 
\be 
r_A^{\prime} \leq - 10^{-3}
\mlab{RPLB}  
\ee
if \ot~is conserved. 
\item 
This would have a dramatic impact on 
$K^{\pm} \to \pi ^{\pm} \pi ^0$ decays. For \mref{RPLB} implies a
sizeable \cpt~asymmetry there 
\be 
\frac
{\Gamma (K^+\to \pi ^+ \pi ^0) - \Gamma (K^-\to \pi ^- \pi ^0)}
{\Gamma (K^+\to \pi ^+ \pi ^0) + \Gamma (K^-\to \pi ^- \pi ^0)}
> 10^{-3}
\ee 
With \cpt~symmetry we predict here 
a direct \cp~asymmetry 
of at most ${\cal O}(10^{-6})$ due to electromagnetic 
corrections. Thirty year old data yield 
$(0.8 \pm 1.2) \cdot 10^{-2}$. Upcoming experiments 
will produce a much better measurement. 
\end{itemize}

%%%%%%%%%%%%%
\subsection{$K_L \to \pi ^+\pi ^- e^+e^-$}
%%%%%%%%%%%%

If the photon polarization 
$\vec \epsilon _{\gamma}$ in 
$
K_L \to \pi ^+ \pi ^- \gamma 
$ 
were measured, we could form the \cp~and \ot~odd 
correlation $P_{\perp}^{\gamma} \equiv 
\langle \vec \epsilon _{\gamma} \cdot 
(\vec p_{\pi ^+} \times \vec p_{\pi ^-})\rangle$. 
A more practical realization of this idea is to analyze 
$K_L \to \pi ^+ \pi ^- e^+ e^-$ which proceeds like 
$
K_L \to \pi ^+ \pi ^- \gamma ^* \to \pi ^+ \pi ^- e^+ e^-
$.  
It allows to determine a \cp~and \ot~odd moment 
$\langle A\rangle$ related to 
$P_{\perp}^{\gamma}$ by measuring the correlation 
between the $\pi ^+ \pi ^-$ and $e^+ e^-$ planes. This effect 
was predicted to be \cite{segal}
\be  
\langle A\rangle = (14.3 \pm 1.3)\%
\mlab{ATH}
\ee 
and observed by 
KTeV \cite{ktev}: 
\be 
\langle A\rangle = (13.6 \pm 2.5\pm 1.2)\% 
\mlab{AEXP}
\ee
It is 
mainly due to the interference between the bremsstrahlung 
process 
$K_L \Rightarrow K_{\cp +} \to \pi ^+ \pi ^- \to 
\pi ^+ \pi ^- \gamma ^*$ and a one-step M1 reaction 
$K_L \to \pi ^+ \pi ^- \gamma ^*$. The former is 
\cp~violating and described by $\eta _{+-}$ 
{\em irrespective} of 
the theory underlying \cp~violation. 

It is a remarkable measurement since it has revealed a 
huge \cp~asymmetry in a rare channel that had not been observed  
before. While \ot~odd correlations have been seen before in 
production processes and in nonleptonic hyperon decays, those 
-- due to their sheer magnitude -- had to be blamed on 
final state interactions; such an explanation turned out 
to be consistent with what we know about those. The 
quantity $\langle A \rangle$ on the other hand is a 
\ot~odd correlation {\em sui generis} since it has a chance 
to be generated by microscopic \ot~violation.

Yet the most intriguing question is 
what does this measurement teach us about 
\ot~violation without reference to \cpt~symmetry? 
The answer is: Nothing really! For we have just shown -- 
by giving a concrete example -- that if we are sufficiently determined 
we can dial \cpt~violation in such a way that both the modulus and 
phase of $\eta _{+-}$ are reproduced even with \ot~invariant dynamics, 
and it is $\eta _{+-}$ that controls $\langle A \rangle$.  

%%%%%%%%%%%%%%%%%%%%%%
\subsubsection{A Comment on the Intricacies of Final State
Interactions}
%%%%%%%%%%%%%%%%%%%%%%
It is well-known that a non-vanishing \ot -odd correlation does
not necessarily establish \ot~violation since final state interactions
can induce it even if \ot~is conserved. Yet even so the reader 
might be surprised by our findings that a 
value of $\langle A\rangle$ 
as large as 
10\% does not establish \ot~violation.  For it 
would be tempting to argue that in the case at hand final state
interactions could not induce an effect even within an order of
magnitude of the observed size. The argument might proceed as follows: 
$\langle A \rangle$ reflects the correlation between the 
$\pi ^+ -\pi ^-$ and the $e^+ - e^-$ planes; their relative
orientation can be affected by final state interactions -- but only of
the electromagnetic variety; then $\langle A \rangle \gg 1\%$ could
not arise. 

If nothing else, our brute force 
scenario shows that such an argument is
fallacious. This can be seen also more directly. 
As stated above there 
are two different contributions to 
$K_L \to \pi ^+ \pi ^- e^+ e^-$, namely the M1 amplitude which is 
\cp~neutral, and the bremsstrahlung one due to the presence of 
\cp~violation. One should note that the presence of the this 
second amplitude requires neither \ot~violation nor final state 
interactions!  

Let us assume for the moment that arg $\eta _{+-} = 0$ were to hold. 
Ignoring final state interactions both in the M1 and the 
bremsstrahlung amplitudes one obtains $\langle A \rangle =0$, 
since the former is imaginary and the latter real now. When the final
state interactions are switched back on, they affect the two
amplitudes differently. Interference can take place, and one finds 
(with arg $\eta _{+-} = 0$) $\langle A \rangle \sim 8 \%$. 
How can the orientation of the 
$\pi ^+ - \pi ^-$ and the 
$e^+ - e^-$ planes get shifted so much by strong final 
state interactions?  
The fallacy of the intuitive argument sketched above derives from its
purely classical nature. In quantum mechanics it is not surprising at
all that phase shifts between coherent amplitudes 
change angular correlations.

%%%%%%%%%%%%%
\section{Summary  
\label{SUMMARY}
}
%%%%%%%%%%%%%%
In this note we have listed the information we 
can infer  
on \ot~and \cpt~invariance from the data on kaon decays. 
Our reasoning was guided by the conviction that once we contemplate
\cpt~breaking the notion of a reasonable or natural assumption 
starts to resemble an oxymoron. 
 
Our findings can be summarized as follows: 
\begin{itemize} 
\item 
The presence of \ot~violation in $\Delta S\neq 0$ dynamics 
has been shown without invoking 
\cpt~symmetry through the Kabir 
test performed by CPLEAR. Yet their analysis had to 
assume semileptonic kaon decays to be \cpt~symmetric or 
it had to impose the Bell-Steinberger relation in its 
conventional form. We do not view either assumption as 
qualitatively more sacrosanct than \cpt~symmetry. 
\item 
$\phi _{+-,00}$ lie within $2^o$ of what is expected 
from \cpt~symmetry.
\item 
A meaningful yardstick for calibrating bounds on 
limitations to \cpt~symmetry is provided by \cp~asymmetries. 
\cpt~breaking forces could -- empirically -- 
still be as large as 
few percent 
of \cp~violating forces. 
\item 
It is grossly misleading to calibrate the bound on 
$M_{\over K} - M_K$ inferred from $\phi _{+-}$, 
$\phi _{00}$ and $\phi _{SW}$ to the kaon mass.  
\item 
The measured values of $\eta _{+-}$ and $\eta _{00}$ provide 
us with little information on the level of 
\ot~versus \cpt~violation.  
More specifically $\eta _{+-}$ -- both its modulus as well as 
its phase -- can be reproduced with \ot~invariant dynamics 
(unless one imposes the Bell-Steinberger relation): 
\begin{itemize}
\item 
This is achieved by carefully adjusting 
\cpt~violation in $\Delta S=1 \& 2$ transitions. 
\item 
The observed level of direct \cp~violation -- 
$\eta _{+-} \neq \eta _{00}$ -- is {\em not} a natural consequence of
such a 
scenario. However it could arise due to a fine-tuning of 
$\delta _2 - \delta _0 - \phi _{SW}$ -- which had to be viewed as 
completely accidental -- or to a compensation of direct \cpt~violation
in $K_L \to [\pi \pi ]_0$ and $K_L \to [\pi \pi ]_2$. 
\item 
In the latter subscenario one is stuck with a \cpt~asymmetry 
in $K^{\pm}\to \pi ^{\pm} \pi ^0$ that could be 
up to few$\times 10^{-3}$ without upsetting any known empirical 
bound. 
\item    
The KTeV observation of a large \cp~and \ot~odd correlation 
in $K_L \to \pi ^+ \pi ^- e^+ e^-$ in agreement with 
theoretical predictions is highly intriguing, 
yet does {\em not} constitute an unequivocal 
signal for \ot~violation. This has also been noted before 
\cite{ELLIS} using a different line of reasoning. 
\end{itemize} 
\item 
We are fully aware that our construction is purely ad-hoc without any
redeeming theoretical feature. Nevertheless we do not view it as 
l'art pour l'art (or more appropriately non-art pour non-art): 
\begin{itemize}
\item 
We have shown by constructing an explicit counter-example that the 
\ot~odd correlation observed in $K_L \to \pi ^+ \pi ^- e^+ e^-$ 
does {\em not} establish \ot~violation without invoking the 
\cpt~theorem. 
\item 
As a by-product we have found that $K^{\pm} \to \pi ^{\pm}\pi ^0$ 
could exhibit a \cpt~asymmetry large enough to become observable
soon. 
\end{itemize}
 
\end{itemize}
Finally we would like to add the remark that even negative 
searches for \cpt~violation in kaon transitions will 
{\em not} free 
us from the obligation to probe for such effects in beauty 
meson decays at the $B$ factories.

%\item 
%Data from the $\Phi$ factory $DA\Phi NE$  at Frascati 
%will probably crucial for achieving further insights. 

%%%%%%%%%%%

\vskip 3mm  
{\bf Acknowledgements} 
\vskip 3mm  
We thank Y. Nagashima for discussions. 
We are grateful to P. Bloch and his colleagues for pointing out 
several relevant errors and omissions in the original version 
of this paper. 
The work of I.I.B. has been supported by the NSF under the grant 
PHY 96-0508 and that of A.I.S. by Grant-in-Aid for Special Project 
Research (Physics of CP violation).

%%%%%%%%%%%%%%%%%%%%%%%%%%%%%%%

\end{document}